\documentclass[manuscript]{aastex}
\usepackage{txfonts}
\usepackage{graphicx}

\slugcomment{Accepted to ApJ}

\shorttitle{Uniform Infall toward the Cometary H II Region in the G34.26+0.15 Complex?} \shortauthors{Liu et al.}

\begin{document}

\title{Uniform Infall toward the Cometary H II Region in the G34.26+0.15 Complex?}
\author{Tie Liu\altaffilmark{1}, Yuefang Wu\altaffilmark{1}, Huawei Zhang\altaffilmark{1}}

\altaffiltext{1}{Department of Astronomy, Peking University, 100871,
Beijing China; liutiepku@gmail.com, ywu@pku.edu.cn }

\begin{abstract}
Gas accretion is a key process in star formation. However, the gas infall detections in high-mass star forming regions with high-spatial resolution observations are rare. Here we report the detection of gas infall towards a cometary ultracompact H{\sc ii} region "C" in G34.26+0.15 complex. The hot core associated with "C" has a mass of $\sim$76 M$_{\sun}$ and a volume density of 1.1$\times$10$^{8}$ cm$^{-3}$. The HCN (3--2), HCO$^{+}$ (1--0) lines observed by single-dishes and the CN (2--1) lines observed by the SMA show redshifted absorption features, indicating gas infall. We found a linear relationship between the line width and optical depth of the CN (2--1) lines. Those transitions with larger optical depth and line width have larger absorption area. However, the infall velocities measured from different lines seem to be constant, indicating the gas infall is uniform. We also investigated the evolution of gas infall in high-mass star forming regions. At stages prior to hot core phase, the typical infall velocity and mass infall rate are $\sim$ 1 km~s$^{-1}$ and $\sim10^{-4}$ M$_{\sun}\cdot$yr$^{-1}$, respectively. While in more evolved regions, the infall velocity and mass infall rates can reach as high as serval km~s$^{-1}$ and $\sim10^{-3}-10^{-2}$ M$_{\sun}\cdot$yr$^{-1}$, respectively. Accelerated infall has been detected towards some hypercompact H{\sc ii} and ultracompact H{\sc ii} regions. However, the acceleration phenomenon becomes inapparent in more evolved ultracompact H{\sc ii} regions (e.g. G34.26+0.15).

\end{abstract}


\section{Introduction}
Our understandings towards high-mass star formation are still sketchy due to the difficulties in observations caused by their large distance, large extinction, clustering forming behaviors as well as the strong interactions between the forming young stars and their surroundings \citep{zin07}. Is high-mass star formation merely a scaled-up version of low-mass star formation with higher accretion rates or does high-mass star formation owe particular properties which are dramatically different from low-mass star formation? There are two most promising models to form high-mass stars (M $>$8 M$_{\sun}$). One is called "monolithic
collapse and disk accretion" \citep{york02,mck03}, and the other "competitive accretion" \citep{bonn02,bonn04}. The former suggests that high-mass stars form directly from isolated massive gas clumps as do low-mass stars, but with a much larger accretion rate. The latter claims that high-mass stars form in the center of a cluster through competing for cloud gas with the other off-center protostars. To distinguish different models, detailed observations of gas accretion in high-mass star forming regions are needed.

So far, several surveys of gas infall have been carried out with single-dishes and gas infall is found to be common in high-mass star forming regions \citep{ful05,wu03,wu07}. However, due to the poor spatial resolution, single-dish observations can not resolve individual cores or protostars. Thus it is hard to investigate the accretion mode in high-mass star formation with single-dish observations. In recent years, high spatial resolution observations of infall in high-mass star forming regions with interferometers have been boomed \citep{solho05,zap08,wu09,gir09,shi10,Liu11a,Liu11b,zhu11,qiu11,qiu12}. Infall was detected in high-mass star forming regions at various evolutionary stages. Towards those at early evolutionary phases (prior to hot core phase), optically thick lines often show blue profile plus an inverse P-cygni profile with a small infall velocity ($\sim$ 1 km~s$^{-1}$) and a shallow absorption dip \citep{zhu11,Liu11a}. However, towards those more evolved regions with bright continuum emission, where the protostars have gained more mass and the UV radiation becomes strong to resist gravity, inverse P-cygni profiles were often detected in lines \citep{zap08,wu09,gir09,shi10,Liu11b,qiu11,qiu12}. The infall velocities (several km~s$^{-1}$) and mass infall rates ($\sim10^{-3}$ M$_{\sun}\cdot$yr$^{-1}$) in these regions are pretty high and the core even collapse faster inside than outside \citep{wu09,qiu11}. However, such high spatial resolution studies are still rare. More samples are needed to constrain the properties of gas infall in high-mass star forming regions.

In this paper, we report the results of a high spatial resolution study with the SMA towards a cometary ultracompact H{\sc ii} region G34.26+0.15.  Locating at a distance of 3.7 kpc \citep{ku94}, G34.26+0.15 is a high-mass star forming complex. Previous centimeter observations \citep{reid85} have detected a shell-like H{\sc ii} region ("D"), a cometary ultracompact H{\sc ii} region ("C"), and two hypercompact H{\sc ii} region ("A" and "B") in G34.26+0.15 complex. A hot core with a kinetic temperature of 160$\pm$30 K was also detected towards the cometary ultracompact H{\sc ii} region "C" \citep{moo07}. In previous works, gas infall towards the hot core associated with the cometary ultracompact H{\sc ii} region "C" was never reported. Here we report the first detection of gas infall in this region with line observations from both single-dish and interferometer observations.

\section{Observations}
The observations of G34.26+0.15 with the IRAM 30 m telescope at Pico Veleta, Spain were carried out in August 2005.
The detail of the observations can be found in \cite{wu07}. In this paper, C$^{34}$S(5--4), C$^{17}$O (1--0) and HCO$^{+}$ (1--0) lines were
used.

The single-pointing observation of HCN (3--2) was carried out in April 2005 with the JCMT telescope in Hawaii. The
JCMT beam size for HCN (3--2) was 18$\arcsec$.3, and the main beam
efficiency was 0.69. During the observation, the weather is pretty good with $\tau$=0.045. The system temperature is 271 K. The noise level
of the line is about 0.15 K.

The SMA data were obtained from SMA raw data archive. The observations of G34.26+0.15 were carried out with the SMA in
April 2011 in its compact configuration.
The phase reference center was R.A.(J2000)~=~18$^{\rm h}$53$^{\rm m}$18.573$^{\rm s}$ and
DEC.(J2000)~=~$01\arcdeg14\arcmin58.3\arcsec$. The 1 receiver 4 GHz mode with uniform spectral resolution of $\sim$0.8125 MHz (128 channels per
chunk) was adopted. The 230 GHz receivers were tuned to 227.5 GHz for the lower sideband (LSB)
and 239.5~GHz for the upper sideband (USB).

In the observations, Saturn
and QSO 3c279 were observed for antenna-based bandpass correction.
QSOs 1741-038 and 1911-201 were employed for antenna-based gain correction.
Neptune was observed for flux-density calibration.

Miriad was employed for calibration and imaging \citep{sau95}. The 1.3 mm continuum data were acquired by
averaging all the line-free channels over both 4 GHz sidebands. MIRIAD task "selfcal" was employed to
perform self-calibration on the continuum data.
The gain solutions from the self-calibration were applied to the
line data. The synthesized beam size and 1 $ \sigma$ rms of the dust emission observed in the compact configuration
is $2\arcsec.90\times2\arcsec.39$ (PA=-47$\arcdeg$.1) and 25 mJy~beam$^{-1}$, respectively.

Spitzer/IRAC data were also retrieved from the database of GLIMPSE \footnote{http://irsa.ipac.caltech.edu/data/SPITZER/GLIMPSE/}.

\section{Results}
\subsection{Overall picture of G34.26+0.15 complex}
G34.26+0.15 is a star forming complex composed of four H{\sc ii} regions ("A", "B", "C" and "D") at different evolutionary stages \citep{reid85}. As shown in Figure 1, the PAH emission (pink contours) revealed by Spitzer/IRAC 8 $\micron$ band forms a shell-like structure, which shows the boundary of the expanding H{\sc ii} region "D". The SCUBA 850 $\micron$ emission (red contours) reveals a compact clump, which harbors two hypercompact H{\sc ii} regions ("A" \& "B") and one ultracompact H{\sc ii} region ("C"). The sequential star formation in this region seems to be induced by the expansion of the H{\sc ii} region "D".

The strong, extended emission in the 4.5 $\micron$ band of Spitzer/IRAC is usually thought to be produced by shock-excited molecular H$_{2}$ and CO in protostellar outflows \citep{nor04,rea06,smi06,dav07,tak10}. To reduce the contamination from the stellar emission, we present the IRAC [4.5]/[3.6] flux ratio image in color in Figure 1. The flux ratio is comparable or higher than $\sim$1.5 in the jets in contrast to the stars (flux ratio $\ll$1.5) \citep{tak10}. As depicted by the long white dashed lines, one can identify several elongated structures (jets?) from the ratio image, especially at the NW and SW of the dust core. If the large [4.5]/[3.6] ratio can reveal the distribution of shocked gas, there seems to exist multiple jets generated from the G34.26+0.15 complex. As shown in Figure 2, one can identify broad line wings from HCN (3--2) line. The terminal velocity of the red wing even reaches as high as 40 km~s$^{-1}$, indicating energetic outflow motions.

\subsection{1.3 mm dust emission obtained from the SMA observations}

As shown in grey scale in Figure 3, the 1.3 mm continuum emission from SMA observations reveals a dense core. The positions of two hypercompact H{\sc ii} regions ("A" \& "B") and one ultracompact H{\sc ii} region ("C") are marked with "stars". The expanding shell-like H{\sc ii} region "D" is located to the southeast of "B", which is not marked in the Figure 3. The emission peak of the 1.3 mm continuum emission coincides with the cometary ultracompact H{\sc ii} region "C". The peak flux of the 1.3 mm emission is 8.31$\pm$0.38 Jy~beam$^{-1}$. The total integrated flux and deconvolved size of the core are 11.41 Jy and 1$\arcsec.74\times1\arcsec.29$ (P.A.=-50.5$\arcdeg$), respectively. Since the integrated flux densities at 1.3 cm and 2.8 mm are as high as 5 Jy and 6.7$\pm$0.4 Jy \citep{moo07}, respectively, the free-free contribution to the 1.3 mm emission can not be ignored. Adopting a spectral index of 0.2 \citep{moo07}, the expected free-free emission at 1.3 mm is 7.8 Jy. Therefore the dust contribution at 1.3 mm is 3.6 Jy.

Assuming that the dust emission is optically thin and the dust temperature equals the kinetic temperature 160 K \citep{moo07},
the total gas mass of the dust core can be obtained with the formula
$M=S_{\nu}D^{2}/\kappa_{\nu}B_{\nu}(T_{d})$, where $S_{\nu}$ is the
flux of the dust emission at 1.3 mm, D is the distance, and $B_{\nu}(T_d)$ is the Planck
function. The dust opacity at
1300 $\micron$ derived from a gas/dust model with thin ice mantles
\citep{oss94} is $\kappa_{\nu}=0.009$ cm$^{2}$g$^{-1}$. Here the ratio of gas to dust is taken as 100. Thus the mass and volume density of the envelope within the inner $\sim$0.01 pc radius are 76 M$_{\sun}$ and 1.1$\times$10$^{8}$ cm$^{-3}$, respectively.

\subsection{Gas infall towards the dust core}
As shown in Figure 2, both of the optically thin C$^{34}$S (5--4) and C$^{17}$O (1--0) lines at the dust core are single-peaked. From gaussian fit, the peak velocities of the C$^{34}$S (5--4) and C$^{17}$O (1--0) lines are 57.9 and 58.2 km~s$^{-1}$, respectively. Here we take the average value of 58.05 km~s$^{-1}$ as the systemic velocity of the dust core. In contrast, both the HCN (3--2) and HCO$^{+}$ (1--0) show a redshifted absorption dip at around 61 km~s$^{-1}$, indicating gas infall towards the dust core.

All the spectra of the CN (2--1) transitions from the SMA observations show redshifted absorption features. Figure 4 presents the spectra of four CN (2--1) transitions, all of which show redshifted absorption. The parameters of the unblended CN (2--1) lines are summarized in Table 1.

The redshifted absorption of the lines (inverse P-cygni profile) is regarded as the evidence of a cold infalling layer in front of a bright continuum background. The redshifted absorption feature of the HCN (3--2), HCO$^{+}$ (1--0) and CN (2--1) lines indicates that gas is still falling down to the dense inner part of the cometary H{\sc ii} region "C".

\section{Discussion}

\subsection{Uniform infall towards the cometary H{\sc ii} region "C"}
Infall motion has been identified towards the hyper/ultra compact H{\sc ii} regions \citep{wu09,qiu11}. The dust/gas cores associated with these H{\sc ii} regions always collapse faster inside than outside \citep{wu09,qiu11}. While towards the cometary H{\sc ii} region "C" in G34.26+0.15 complex, the HCN (3--2) and HCO$^{+}$ (1-0) from single-dish observations and the CN (2--1) lines from the SMA observations all show a redshifted absorption dip around 61 km~s$^{-1}$, which is $\sim$ 3 km~s$^{-1}$ away from the systemic velocity. This indicates that the infall motion is uniform at different spatial scale.

The critical densities $n_{crit}$ of different transitions of CN (2--1) lines can be calculated as:
\begin{equation}
n_{crit}=\frac{A_{ij}}{K_{ij}}
\end{equation}
where $A_{ij}$ and $K_{ij}$ are the Einstein coefficients and collisional rate coefficients. We adopted the collisional rate coefficients at 100 K from LAMDA \footnote{http://home.strw.leidenuniv.nl/~moldata/} in calculations. The calculated critical densities are listed in the fifth column of Table 1. The critical densities of CN (2--1) lines varies from 7.4$\times10^{5}$ to 1.4$\times10^{7}$ cm$^{-3}$, indicating the various transitions of CN (2--1) can be used to trace different layers of the gas core. In Figure 3, we plot the contours of the absorption of various CN (2--1) transitions. One can find that various transitions of CN (2--1) do trace different layers of the gas core. The optical depth of each CN (2--1) line can be estimated from:
\begin{equation}
\tau_{L}=-\textrm{ln}(\frac{I_{L}}{I_{C}})=-\textrm{ln}(1+\frac{\Delta I_{L}}{I_{C}})
\end{equation}
where $\Delta I_{L}=I_{L}-I_{C}$ is the observed line intensity at the continuum peak and the $I_{C}$ is
the observed peak continuum intensity. The optical depth is listed in the last column of Table 1. From Figure 3, we noticed that the transitions of CN (2--1) with larger optical depth also have larger absorption area.

The line widths of various CN (2--1) transitions are listed in the 8th column of Table 1. We found that the transitions with smaller line widths also have smaller absorption area in Figure 3. This phenomenon is quite similar to the "Larson relationship" found in molecular clouds \citep{lar81}. It seems that the non-thermal motion in the outer part of the core is more active than that in the inner part. As shown in Figure 5, there exists a linear relationship between the line width and optical depth of CN (2--1) lines. Thus the optical depth and line width can be used to distinguish different layers of the core. The infall velocities were measured from various CN (2--1) transitions with $V_{in}=V_{obs}-V_{sys}$. The absorption velocity $V_{obs}$ of each line was obtained from gaussian fit and listed in the 6th column of Table 1. While the infall velocities are listed in the 7th column. As shown in Figure 5, The infall velocity seems to be constant for various optical depth and line width, indicating the infall motion is uniform in this region. The average infall velocity measured from CN (2--1) lines is 3.25 km~s$^{-1}$.

\subsection{The evolution of gas infall in high-mass star forming regions}
In table 2, we summarized the parameters of 11 high-mass star forming regions with infall detections by (sub)millimeter interferometer observations. The infall velocities were measured with $V_{in}=V_{obs}-V_{sys}$. The mass accretion rates, $\dot{M}_{in}$, can be estimated using the simple expression:
\begin{equation}
\dot{M}_{in}=4\pi r_{in}^{2}\mu m_{H}nV_{in}
\end{equation}
where $\mu$ is the mean molecular weight and $m_{H}$ is the mass of hydrogen. Assuming that the infall
speeds $V_{in}$ arise from the velocity gain of gas free-falling from rest at $r=\infty$ to $r_{in}$, thus $r_{in}$
can be calculated as:
\begin{equation}
r_{in}=\frac{2GM}{V_{in}^{2}}
\end{equation}
The mean volume density n within $R_{in}$ is estimated as:
\begin{equation}
n=\frac{M}{4/3\cdot\pi r_{in}^{3}\mu m_{H}}
\end{equation}

The calculated mass accretion rates are listed the fifth column of Table 2. We also summarized the parameters of two low-mass star forming regions for comparison.
From Table 2, one can find that the infall velocities and mass accretion rates in high-mass star forming regions are much larger than those in low-mass star forming regions. The typical infall velocity and mass accretion rate in low-mass star forming regions are $\sim$0.5 km~s$^{-1}$ and $\sim10^{-5}$ M$_{\sun}\cdot$yr$^{-1}$, respectively. While in high-mass star forming regions, the infall velocity and mass accretion rate can reach as high as several km~s$^{-1}$ and $\sim10^{-3}$ M$_{\sun}\cdot$yr$^{-1}$, respectively.

However, the infall velocity and mass accretion rate change with time in high-mass star forming regions. At evolutionary stages earlier than hot core phase (e.g. W3SE-SMA1 and JCMT 18354-0649S), the typical infall velocity and mass accretion rate are $\sim$ 1 km~s$^{-1}$ and $\sim10^{-4}$ M$_{\sun}\cdot$yr$^{-1}$, respectively. At hot core phase (including hypercompact and early phases of ultracompact H{\sc ii} regions), the infall velocity and mass infall rates can reach as high as serval km~s$^{-1}$ and $\sim10^{-3}-10^{-2}$ M$_{\sun}\cdot$yr$^{-1}$, respectively. But even more amazing is that during this stage accelerated gas infall was often revealed (e.g. NGC 7538 IRS 1, G10.6-0.4 and G19.61-0.23). In other words, with different molecular tracers, the infall velocities inferred in the inner layers are larger than the infall velocities in the outer layers. However, the infall acceleration phenomenon becomes inapparent in more evolved ultracompact H{\sc ii} regions (e.g. G34.26+0.15 "C").

In Figure 6, we found a tight relationship between the infall velocity and the total dust/gas mass. The relationship can be depicted with a power-law equation. If taking the low-mass star forming regions into account, the power index is 0.34. While the power index becomes 0.36 if we only consider the high-mass star forming regions. This tight relationship indicates that gravity plays the dominated role in gas accretion.

Whether the core collapses depends on the balance of gravity, thermal and turbulent support, magnetic filed and so on. As the protostars in the cores evolve, the internal heating and UV illumination may also play an important role in resisting gravity. More detailed observations and numerical simulations are needed to address how these factors function in the gas accretion as the protostars evolve.

\section{Summary}
Here we report the detection of gas infall towards of cometary ultracompact H{\sc ii} region "C" in G34.26+0.15 complex. From the 1.3 mm continuum emission, we estimated the total dust/gas of the hot core associated with the cometary ultracompact H{\sc ii} region "C" is about 76 M$_{\sun}$. Both the HCN (3--2) line observed by JCMT and the HCO$^{+}$ (1--0) observed by IRAM show a redshifted absorption dip at around 61 km~s$^{-1}$, indicating gas infall in this region. We also detected absorption lines of multiple CN (2--1) transitions from the SMA observations. We found a linear relationship between the line width and optical depth of the CN (2--1) lines. Those transitions with larger optical depth and line width have larger absorption area. However, the infall velocities measured from different lines seem to be constant, indicating the gas infall is uniform. The uniform infall may be due to the expansion of the cometary H{\sc ii} region and the feedback from the central protostars in the form of UV radiation and stellar wind.

We also investigated the evolution of gas infall in high-mass star forming regions with the data collected from literatures. We found that the infall velocity and mass infall rate in high-mass star forming regions are much higher than those in low-mass star forming regions. However, the infall also evolves with time. At stages prior to hot core phase, the typical infall velocity and mass infall rate are $\sim$ 1 km~s$^{-1}$ and $\sim10^{-4}$ M$_{\sun}\cdot$yr$^{-1}$, respectively. While in more evolved regions, the infall velocity and mass infall rates can reach as high as serval km~s$^{-1}$ and $\sim10^{-3}-10^{-2}$ M$_{\sun}\cdot$yr$^{-1}$, respectively. Accelerated infall has been detected towards several hypercompact H{\sc ii} and ultracompact H{\sc ii} regions. However, the acceleration phenomenon becomes inapparent in more evolved ultracompact H{\sc ii} regions (e.g. G34.26+0.15). As the protostars in the cores evolve, the internal heating and UV illumination may play an important role in resisting gravity. Thus gas infall in more evolved H{\sc ii} regions may be decelerated or even halted eventually.

\section*{Acknowledgment}
\begin{acknowledgements}
We are grateful to the SMA staff. This work was funded by China Ministry of Science and Technology under State Key Development Program for Basic Research 2012CB821800.
\end{acknowledgements}

\begin{figure}[!bht]
\begin{center}
\begin{tabular}{c}
\includegraphics[scale=0.4,angle=-90]{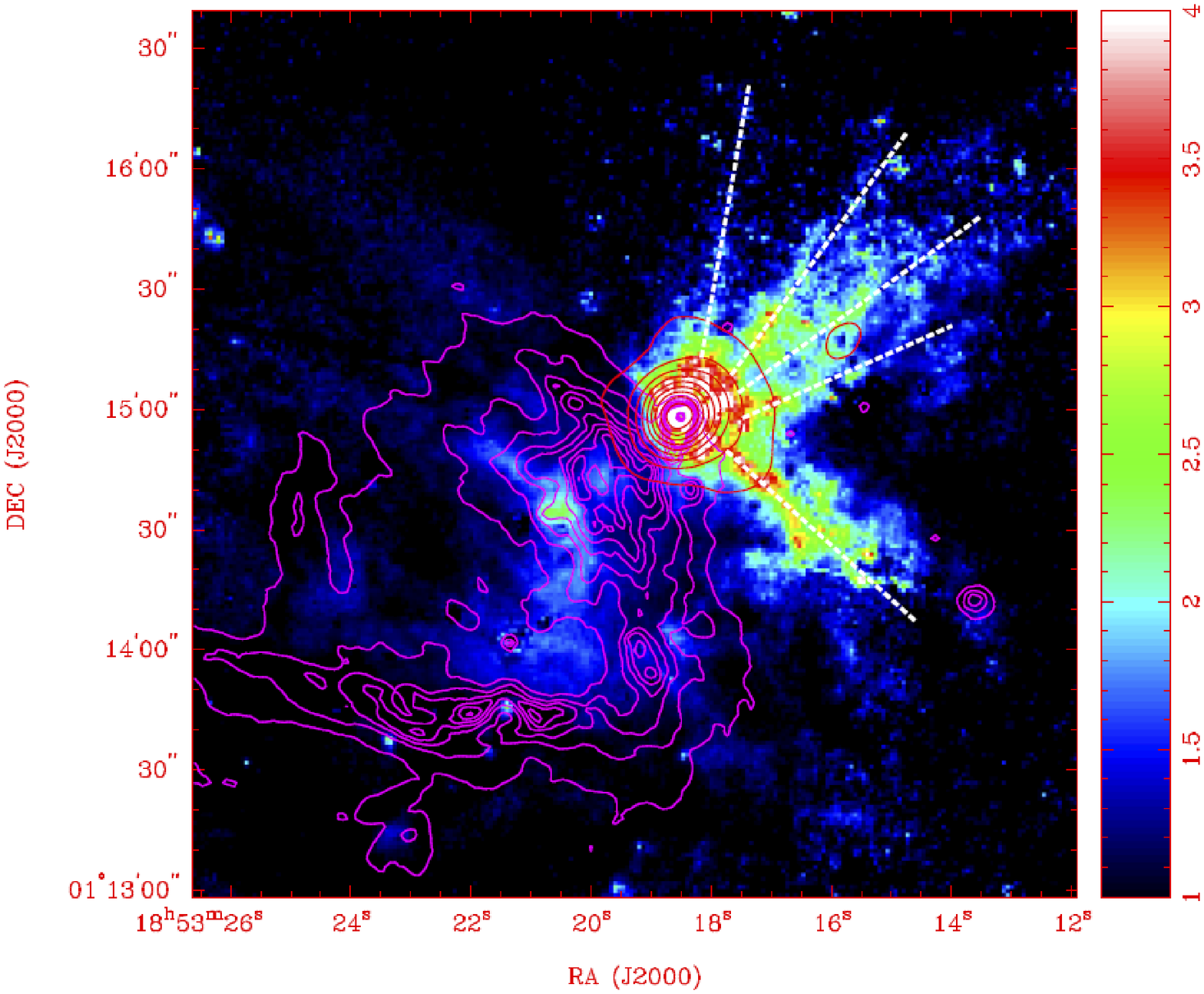}
\end{tabular}
\end{center}
\caption{The flux ratio map of spitzer [4.5]/[3.6] is shown in color scale. The pink contours represent the distribution of 8 $\mu m$ emission (PAH). The contour levels are from 10\% to 90\% in a step of 10\% of the peak emission. The 850 $\mu m$ continuum from SCUBA/JCMT is shown as red solid contours. The contour levels are from 10\% to 90\% in a step of 10\% of the peak emission. The long white dashed lines represent the possible jets/outflows directions.}
\end{figure}

\begin{figure}[!bht]
\begin{center}
\begin{tabular}{c}
\includegraphics[scale=0.4,angle=-90]{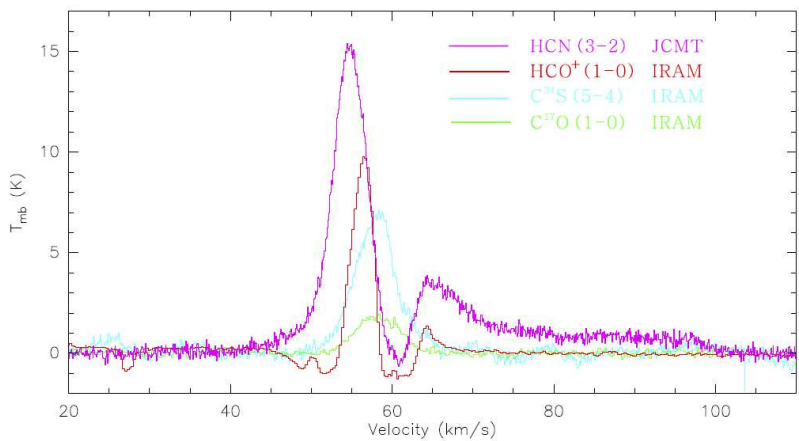}
\end{tabular}
\end{center}
\caption{Single-dish spectra at the dust emission peak of G34.26+0.15.}
\end{figure}

\begin{figure}[!bht]
\begin{center}
\begin{tabular}{c}
\includegraphics[scale=0.4,angle=-90]{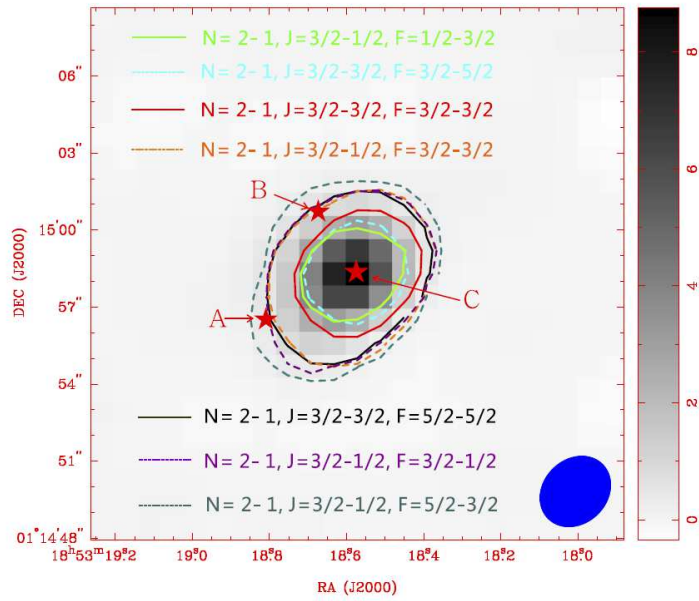}
\end{tabular}
\end{center}
\caption{ CN (2--1) absorption contours overalyed on 1.3 mm continuum emission observed by the SMA. The contour level for each transition is 0.8 Jy~beam$^{-1}$~km~s$^{-1}$. The "stars" mark the positions of two hypercompact and one ultracompact H{\sc ii} regions.}
\end{figure}

\begin{figure}[!bht]
\begin{center}
\begin{tabular}{c}
\includegraphics[scale=0.4,angle=-90]{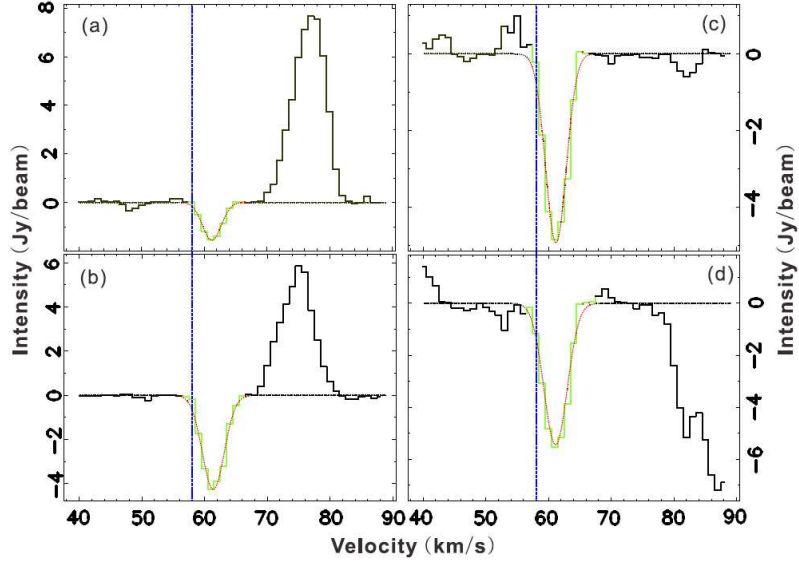}
\end{tabular}
\end{center}
\caption{CN (2-1) absorption lines (highlighted by green color) at the 1.3 mm continuum emission peak of G34.26+0.15. The red lines are the gaussian fits. The systemic velocity (58 km/s) of G34.26+0.15 is marked by blue dashed lines. (a).N=2-1, J=3/2-3/2, F=3/2-3/2 (b). N=2-1, J=3/2-3/2, F=5/2-5/2 (c).N=2-1, J=3/2-1/2, F=3/2-3/2 (d).N=2-1, J=3/2-1/2, F=3/2-1/2. }
\end{figure}

\begin{figure}[!bht]
\begin{center}
\begin{tabular}{c}
\includegraphics[scale=0.4,angle=0]{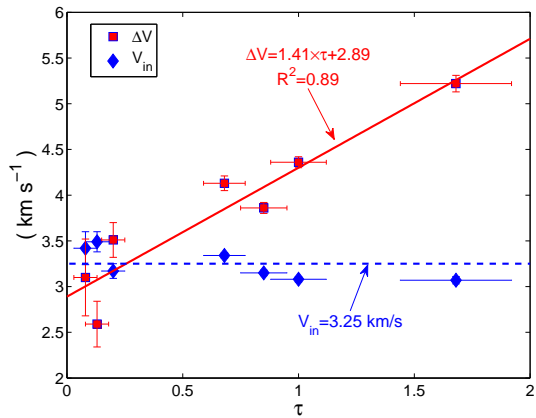}
\end{tabular}
\end{center}
\caption{Infall velocity and line width versus optical depth of CN (2--1) lines.}
\end{figure}

\begin{figure}[!bht]
\begin{center}
\begin{tabular}{c}
\includegraphics[scale=0.4,angle=-90]{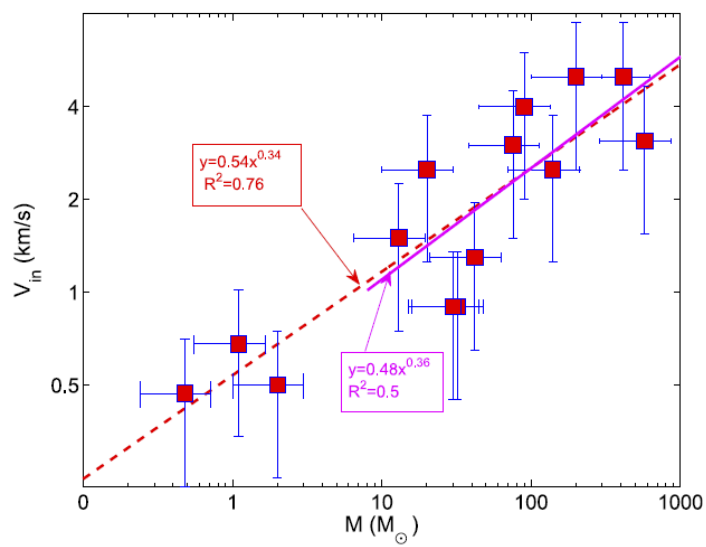}
\end{tabular}
\end{center}
\caption{The relationship between the infall velocity and the total dust/gas mass. The red dashed line represents the linear
fit to both the high-mass and low-mass data points. While the pink solid line shows the linear fit only to the high-mass data points. We assume 50\% error in the measurement of both infall velocity and mass.}
\end{figure}

\clearpage

\begin{deluxetable}{ccccccccccccccc}
\tabletypesize{\scriptsize} \tablecolumns{15} \tablewidth{0pc}
\tablecaption{Parameters of CN absorption lines} \tablehead{
 \colhead{Transition} & \colhead{Frequency}
&\colhead{Log10 (A$_{ij}$)} & E$_{u}$ & n$_{crit}$ & V$_{obs}$ & V$_{in}$ & $\Delta$V &$\tau$ \\
\colhead{} & (GHz) & \colhead{ (Log10 (s$^{-1}$))} &
\colhead{(K)} & (10$^{6}$ cm$^{-3}$) & (km~s$^{-1}$) & (km~s$^{-1}$) & (km~s$^{-1}$) &    } \startdata
N=2-1, J=3/2-3/2, F=3/2-3/2 & 226.315   & -5.00   & 16.31 & 1.62   & 61.22(0.08)  & 3.17(0.08)  &   3.51(0.19)  & 0.20(0.05) \\
N=2-1, J=3/2-3/2, F=3/2-5/2 & 226.333   & -5.34   & 16.31 & 0.74   & 61.54(0.11)  & 3.49(0.11)  &   2.59(0.25)  & 0.13(0.05) \\
N=2-1, J=3/2-3/2, F=5/2-5/2 & 226.360   & -4.79   & 16.31 & 2.63   & 61.39(0.03)  & 3.34(0.03)  &   4.13(0.08)  & 0.68(0.09) \\
N=2-1, J=3/2-1/2, F=1/2-3/2 & 226.617   & -4.97   & 16.31 & 1.55   & 61.47(0.18)  & 3.42(0.18)  &   3.10(0.42)  & 0.08(0.05) \\
N=2-1, J=3/2-1/2, F=3/2-3/2 & 226.632   & -4.37   & 16.31 & 6.15   & 61.20(0.03)  & 3.15(0.03)  &   3.86(0.06)  & 0.85(0.10) \\
N=2-1, J=3/2-1/2, F=5/2-3/2 & 226.660   & -4.02   & 16.31 & 13.66  & 61.12(0.04)  & 3.07(0.04)  &   5.22(0.09)  & 1.68(0.24) \\
N=2-1, J=3/2-1/2, F=3/2-1/2 & 226.679   & -4.28   & 16.31 & 7.60   & 61.13(0.03)  & 3.08(0.03)  &   4.36(0.06)  & 1.00(0.12) \\
\enddata
\end{deluxetable}

\begin{deluxetable}{ccccccccccccccc}
\tabletypesize{\scriptsize} \tablecolumns{15} \tablewidth{0pc}\setlength{\tabcolsep}{0.05in}
\tablecaption{Parameters of High-mass star forming regions with infall detections by (sub)millimeter interferometers. } \tablehead{
 \colhead{Name} & \colhead{Distance}
&\colhead{M} & V$_{in}$\tablenotemark{a}  & $\dot{M}$\tablenotemark{b} &Lines &Stages &Refs\tablenotemark{c} \\
\colhead{}  & \colhead{(kpc)} &
\colhead{(M$_{\sun}$)} & (km~s$^{-1}$) & \colhead{(10$^{-3}$ M$_{\sun}\cdot$yr$^{-1}$)} &  &    } \startdata
\multicolumn{8}{c}{Low-mass star forming regions}\\
\hline
IRAS 16293-2422B    &  0.12  &        2     &    0.49   & 0.04    &     CH$_{3}$OCHO-E ($17_{4,13}-16_{4,12}$)                &  Class0    & 1      \\
                    &        &              &    0.49   & 0.04    &     CH$_{3}$OCHO-A ($17_{4,13}-16_{4,12}$)               &                     & 1      \\
                    &        &              &    0.51   & 0.05    &     H$_{2}$CCO ($11_{1,11}-10_{1,10}$)                    &                     & 1     \\
                    &        &              &    0.7    & 0.12    &     CH$_{3}$OH (9$_{3,6}-8_{2,7}$)                   &                     & 2     \\
NGC 1333 IRAS 4A    & 0.35   &        1.1   &    0.68   & 0.11    &     H$_{2}$CO ($3_{1,2}-2_{1,1}$) &   Class0               & 3      \\
NGC 1333 IRAS 4B    & 0.35   &        0.48  &    0.47   & 0.04    &     H$_{2}$CO ($3_{1,2}-2_{1,1}$) &    Class0                 & 3      \\
\hline
\multicolumn{8}{c}{High-mass star forming regions}\\
\hline                                                                                                                             \\
W3SE-SMA1           &  2     &        32    &     0.9   & 0.26    &      HCN (3-2)                    &    HMPO?            & 4     \\
JCMT 18354-0649S    &  5.7   &        42    &     1.3   & 0.78    &      HCN (3-2)                    &    HMPO             & 5       \\
W51e2-E             &  5.1   &        140   &     2.5   & 5.57    &      HCN (4-3)                    &    HMC?             & 6      \\
G31.41+0.31         &  7.0   &        577   &     3.1   & 10.62   &      C$^{34}$S (7-6)              &    HMC              & 7  \\
W51 IRS2            &  7     &        90    &     4     & 22.81   &      CN (2-1)                     &    HMC              & 8      \\
IRAS 18360-0537 MM1 &  6.3   &        13    &     1.5   & 1.20    &      CN (2-1)                     &    HMC              & 9      \\
NGC 7538 IRS 1      &  2.65  &        20.1  &     2     & 2.85    &      C$^{18}$O (2-1)              &    HC H{\sc ii}     & 10       \\
                    &        &              &     2     & 2.85    &      $^{13}$CO (2-1)              &                     &                  \\
                    &        &              &     3     & 9.62    &      SO ($5_{6}-4_{5}$)           &                     &                  \\
                    &        &              &     4     &         &      HNCO (10$_{0,10}-9_{0,9}$)   &                     &                  \\
                    &        &              &     6     &         &      CH3OH (8$_{-1,8}-7_{0,7}$)   &                     &                  \\
G10.6-0.4           &  6     &        200   &     4     & 22.81   &      HCN (3-2)                    &     HC H{\sc ii}   &  11        \\
                    &        &              &     6     &         &      NH$_{3}$ (3,3)                    &                     &  12     \\
G9.62+0.19 E        &  5.7   &        30    &     0.9   & 0.26    &      CS (7-6)                     &     HC H{\sc ii}   &  13     \\
G19.61-0.23         &  12.6  &        415   &     3.5   & 15.28   &      $^{13}$CO (3-2)              &     UC H{\sc ii}           &  14   \\
                    &        &              &     6     &         &      CN (3-2)                     &                     &                  \\
G34.26+0.15         &  3.7   &        76    &     3     & 12.23    &      CN (2-1),HCN (3-2)           &     cometary UC H{\sc ii}  &  15      \\
\enddata
\tablenotetext{a}{The infall velocities were measured with $V_{in}=V_{obs}-V_{sys}$, which may slightly different from the values in literatures.}
\tablenotetext{b}{only the infall velocities inferred from those molecular lines with absorption area comparable to the continuum emission are used in calculations. }
\tablenotetext{c}{1. \cite{pin12} 2. \cite{zap13} 3. \cite{fra02} 4. \cite{zhu11} 5. \cite{Liu11a} 6. \cite{shi10} 7. \cite{gir09} 8. \cite{zap08} 9. \cite{qiu12}
10. \cite{qiu11} 11. Liu et al.2013, submitted to MNRAS 12. \cite{solho05} 13. \cite{Liu11b} 14. \cite{wu09} 15. this work}
\end{deluxetable}


\begin{thebibliography}

\bibitem[Bonnell et al.(2002)]{bonn02}Bonnell, I. A., Bate, M. R., Clarke, C. J., Pringle, J. E., 2002, \mnras, 323, 785

\bibitem[Bonnell, Vine \& Bate(2004)]{bonn04}Bonnell, I. A., Vine, S. G., \& Bate, M. R., 2004, \mnras, 349, 735

\bibitem[Davis et al.(2007)]{dav07}Davis, C. J., Kumar, M. S. N., Sandell, G., Froebrich, D., Smith, M. D., \& Currie, M. J. 2007, \mnras, 374, 29

\bibitem[Di Francesco et al.(2002)]{fra02}Di Francesco, J., Myers, P. C., Wilner, D. J., Ohashi, N., Mardones, D., 2001, \apj, 562, 770

\bibitem[Fuller et al.(2005)]{ful05}Fuller, G. A., Williams, S. J., Sridharan, T. K., 2005, \aap, 442, 949

\bibitem[Girart et al.(2009)]{gir09}Girart, J. M., Beltr\'{a}n, M. T., Zhang, Q, Rao, R., Estalella, R., 2009, Science, 324, 1408

\bibitem[Kuchar \& Bania(1994)]{ku94}Kuchar, T. A., \& Bania, T. M. 1994, \apj, 436, 117

\bibitem[Larson (1981)]{lar81}Larson, R. B., 1981, \mnras, 194, 809

\bibitem[Liu et al.(2011a)]{Liu11a}Liu, T., Wu, Y., Zhang, Q., Ren, Z., Guan, X., et al., 2011a, \apj, 728, 91

\bibitem[Liu et al.(2011b)]{Liu11b}Liu, T., Wu, Y., Liu, S.-Y., Qin, S.-L., Su, Y.-N., et al., 2011b, \apj, 730, 102

\bibitem[McKee \& Tan(2003)]{mck03}McKee, C. F., \& Tan, J. C., 2003, \apj, 585, 850

\bibitem[Mookerjea et al.(2007)]{moo07}Mookerjea, B., Casper, E., Mundy, L. G., Looney, L. W., 2007, \apj, 659, 447

\bibitem[Noriega-Crespo et al.(2004)]{nor04}Noriega-Crespo, A., et al. 2004, \apjs, 154, 352

\bibitem[Ossenkopf \& Henning(1994)]{oss94}Ossenkopf, V., \& Henning, T. 1994, \aap, 291, 943

\bibitem[Pineda et al.(2012)]{pin12}Pineda, J. E., Maury, A. J., Fuller, G. A., et al., 2012, \aap, 544, L7

\bibitem[Qiu, Zhang \& Menten(2011)]{qiu11}Qiu, K., Zhang, Q., Menten, K. M., 2011, \apj, 728, 6

\bibitem[Qiu et al.(2012)]{qiu12}Qiu, K., Zhang, Q., Beuther, H., Fallscheer, C., 2012, \apj, 756, 170

\bibitem[Reach et al.(2006)]{rea06}Reach, W. T., et al. 2006, \aj, 131, 1479

\bibitem[Reid \& Ho(1985)]{reid85}Reid, M. J., \& Ho, P. T. P. 1985, \apj, 288, L17

\bibitem[Sault et al.(1995)]{sau95}Sault, R. J., Teuben, P. J., \& Wright, M. C. H. 1995, in ASP Conf. Ser. 77, Astronomical Data Analysis Software and Systems IV, ed. R. A. Shaw,
H. E. Payne, \& J. J. E. Hayes (San Francisco, CA: ASP), 433

\bibitem[Shi, Zhao \& Han(2010)]{shi10}Shi, H., Zhao, J.-H., Han, J. L., 2010, \apj, 710, 843

\bibitem[Smith et al.(2006)]{smi06}Smith, H. A., Hora, J. L., Marengo, M., \& Pipher, J. L. 2006, \apj, 645, 1264

\bibitem[Sollins \& Ho(2005)]{solho05}Sollins, P. K. \& Ho, P. T. P. 2005, \apj, 630, 987

\bibitem[Takami et al.(2010)]{tak10}Takami, M., Karr, J. L., Koh, H., Chen, H.-H., Lee, H.-T., 2010, \apj, 720, 155

\bibitem[Wu \& Evans.(2003)]{wu03}Wu, J. \& Evans, N. J., II., 2003, \apj, 592, L79

\bibitem[Wu et al.(2007)]{wu07}Wu, Y., Henkel, C., Xue, R., Guan, X., Miller, M., 2007, \apj, 669, L37

\bibitem[Wu et al.(2009)]{wu09}Wu, Y., Qin, S.-L., Guan, X., Xue, R., Ren, Z., et al., 2009, \apj, 697, L116

\bibitem[York \& Sonnhalter(2002)]{york02} Yorke, H. W. \& Sonnhalter, C., 2002, \apj, 569, 846.

\bibitem[Zapata et al.(2008)]{zap08}Zapata, L. A., Palau, A., Ho, P. T. P., Schilke, P., Garrod, R. T., Rodr¨ªguez, L. F., Menten, K., 2008, \aap, 479, L25	

\bibitem[Zapata et al.(2013)]{zap13}Zapata, L. A., Loinard, L., Rodr¨ªguez, L. F., 2013, \apj, 764, L14

\bibitem[Zinnecker \& Yorke(2007)]{zin07}Zinnecker, H., \& Yorke, H. W., 2007, \araa, 45, 481

\bibitem[Zhu, Zhao \& Wright(2011)]{zhu11}Zhu, L., Zhao, J.-H., Wright, M. C. H., 2011, \apj, 740, 114

\end{thebibliography}
\end{document}